\documentclass[aps,prl,showpacs,twocolumn,amssymb]{revtex4}

\usepackage{blindtext}
\usepackage{graphicx}
\usepackage{bm}

\def\De {\Delta}
\def\Ga {\Gamma}
\def\ep {\epsilon}

\def\e2 {\epsilon-\epsilon_k}
\def\be {\begin{equation}}
\def\ee {\end{equation}}
\def\bea {\begin{eqnarray}}
\def\eea {\end{eqnarray}}
\def\ga {\gamma}
\def\om {\omega}
\def\al {\alpha}

\begin{document}

\centerline{Published in Europhysics Letters, {\bf 112}, 67001 (2015)}


\title{Fermi liquid behaviour in weakly disordered
metals close to a quantum critical point }

\author{George Kastrinakis }

\affiliation{                    
   Institute of Electronic Structure and Laser (IESL), 
Foundation for Research and Technology - Hellas (FORTH), 
P.O. Box 1527, Iraklio, Crete 71110, Greece$^*$ }

\date{6 November 2015}

\begin{abstract} 
We calculate analytically the low temperature quasi-particle scattering 
rate, the conductivity, and the specific heat in weakly disordered metals 
close to a quantum critical point, via the use of a proper fluctuation 
potential $V(q,\omega)$ between the quasi-particles. We obtain typical 
Fermi liquid
results proportional to $T^2$ and $T$ respectively, with prefactors 
which diverge as power laws of the control parameter $a$ upon approaching 
the critical point. The Kadowaki-Woods ratio 
is shown to be independent of $a$ (possibly times a logarithmic dependence
on $a$) only for the
case of three-dimensional ferromagnetic fluctuations. 
Our results are consistent with experiments on the eight materials 
CeCoIn$_5$, Sr$_3$Ru$_2$O$_7$,
YbRh$_2$Si$_2$, La$_{2-x}$Ce$_x$CuO$_4$, Tl$_2$Ba$_2$CuO$_{6+x}$,
CeAuSb$_2$, YbAlB$_4$, and CeRuSi$_2$. 

\end{abstract}


\maketitle

{\bf Introduction.}
Quantum phase transitions take place at zero temperature and are due 
to the zero point quantum fluctuations. 
These fluctuations around the quantum critical point (QCP)
display scale invariance both in space and in time, and result 
in the influence of the QCP over a {\em finite range} of temperature $T$.
Hence the effect of quantum criticality is detectable {\em without} 
actually reaching absolute zero $T$. Typically, some observables
display diverging behaviour upon approaching the QCP.

In itinerant electron systems,
the criticality parameter, which determines the proximity to the respective 
QCP, may depend on the electron filling factor, the pressure,
or the magnetic field (which is related to filling, through the Zeeman
term). A related review can be found in ref. \cite{revi}.
Often these systems display {\em non} Fermi liquid behaviour, in the sense 
that e.g. the temperature dependence of various quantities measured differs
from the standard Fermi liquid (FL) one \cite{revi}.

Our work is motivated by a number of experiments on eight different materials 
\cite{pag,bia,gri,geg1,geg,but,shib,bali,naka,rost,flou}, which display 
typical FL behaviour for appropriately low $T$. That is,
quadratic in $T$ resistivity and linear in $T$ specific heat.
However, the prefactors of these quantities appear to {\em diverge} as the
respective QCP's are approached. We show, via analytic diagrammatic 
calculations, that these facts can be consistently understood as arising
from the exchange of relevant fluctuations among the quasi-particles.
Our approach assumes that we deal with weakly disordered metallic
systems.

\vspace{0.3cm} 
{\bf The model.} We consider the Green's function 
\be
G^{R,A}(k,\ep)=\frac{1}{\ep-\ep_k+\ep_F \pm i /2\tau_o} \;\; , \;\;
\label{gr0}
\ee
with $\ep_k$ the quasiparticle dispersion, $\ep_F$ the Fermi energy and 
$\tau_o$ the momentum relaxation time due to impurities. In the 
weak disorder regime \cite{agd,lee} $\ep_F \tau_o \gg 1$.
$\tau_o^{-1}$ is {\em important as a regulator} in our calculations.
In fact, the characteristic FL $T^2$ dependence of Im $\Sigma$ in 
eqs. (\ref{sig3d}), (\ref{sig2d}) is due to the finite $\tau_o^{-1}$.

The dominant electron-electron interaction is assumed to be
the fluctuation potential (or fluctuation propagator)
\cite{her,mil1,millis,kim,chu,maslov}
\be
V(q,\om)=\frac{g}{-i \om /(D q^2+r)+  \xi^2 (q-q_0)^2 + a} \;\;,
\ee
with $g$ the coupling constant, $\xi$ the 
correlation length and $a$ measuring the distance from 
the QCP. The criticality parameter $a$ depends 
on e.g. the magnetic field $H$, as in the systems of interest mentioned 
below, like $a=h^s$, $h=|H/H_c-1|$, $s>0$, where $H_c$ is the critical field.
The combination $(D q^2 +r)$ in $V(q,\om)$ is considered here for the 
first time. The factor $D q^2$ indicates disorder induced diffusion of the 
quasiparticles, with diffusion coefficient $D$ \cite{lee,gkdef}.
This factor could also originate from antiferromagnetic damping, but 
this origin would not be consistent with our treatment of {\em ferromagnetic} 
fluctuations - c.f. below.
The factor $r$ may originate from the inadvertent presence of elastic magnetic 
impurities in the samples \cite{gkdef}, and should be equal to the relevant 
scattering rate $\tau_S^{-1}$. It can also originate from the
fermiology of a clean system (without disorder). 
$q_0$=0 corresponds to ferromagnetic (FM) fluctuations, while 
finite $q_0$ to antiferromagnetic (AFM) fluctuations.

In general, $\xi$ and $a$ are expected to be related through an equation
of the type \cite{revi,mil1} $\xi^{-2} = d_0 \; a + \beta \; T^w $,
with $d_0, \beta,w$ constants. Below we will assume the Gaussian regime
\cite{her,mil1,millis,kim,chu,maslov},
\be
\xi^{-2} = d_0 \; a   \;\; , \;\;  \label{gaus}
\ee
with $\beta=0$. However, for the purpose of our calculations, we will 
treat $\xi$ and $a$ as independent parameters, and consider the Gaussian 
regime relation 
in the final results. This procedure is entirely consistent, as can be 
seen from the details of the calculations below.

\vspace{0.3cm} 
{\bf Calculation of the scattering rate.}
We calculate the quasi-particle scattering rate as a function 
of $T \rightarrow 0$.
For the self-energy $\Sigma$ we use the relation \cite{agd} (c.f. pg. 183)
\bea
\mbox{Im} \; \Sigma^R(k,\ep) = \sum_{q} \int_{-\infty}^{\infty}d\om \;
\mbox{Im} \; G^R(k-q,\ep-\om) \; \nonumber \\
 \mbox{Im} \; V^R(q,\om) \;
\{\coth(\om/2T) \; + \; \tanh((\ep-\om)/2T) \}\;\;,
\eea
in order to calculate the scattering rate, which equals twice Im $\Sigma$
- c.f. fig 1(a) for the corresponding Feynman diagram. 

\begin{figure}[tb]
  \includegraphics[width=5truecm]{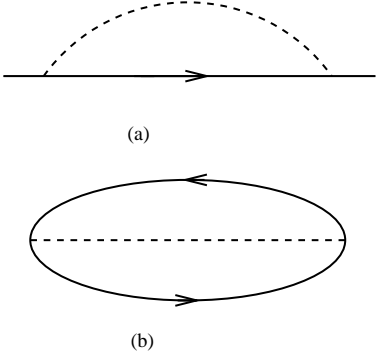}
\caption{Diagrams (a) for the self-energy and (b) for the free energy. 
The continuous
lines are the fermion propagators, i.e. the Green's function of 
eq. (\ref{gr0}), and the dashed
line is the fluctuation mediated interaction $V(q,\om)$.}

\label{fig.1}
\end{figure}

We present the calculation of Im $\Sigma(k,\ep \rightarrow 0)$. 
In 3-D the integration over $q$ is 
$\sum_q = 1/(2\pi)^2 \; \int dq \; q^2 \; \int_{-1}^1 dx $, where 
$x=\cos \theta$, and $\theta$ is the angle between the vectors ${\bf k}$ 
and ${\bf q}$. 
From the factor Im $G(k-q,\ep-\om)$, we have the integral over $x$
\bea
R(q,\omega) = \int_{-1}^1 dx \; \mbox{Im} \; G(k-q,\ep-\om)
=\int_{-1}^1 \frac{dx \; s_o}{s_o^2+(c+b x)^2}  \nonumber \\
=\frac{1}{b } \left[ 
\arctan \left(\frac{c+b}{s_o}\right) - \arctan \left(\frac{c-b}{s_o}\right)
\right] \;\; , \;\;
\eea
where $s_o=1/2\tau_o$, $c=\ep-\omega-\ep_k-q^2/2m+\ep_F$ and $b=k q/m$. 
To be specific, we assume a parabolic dispersion relation $\ep_k=k^2/2m$.
However, the precise form of the dispersion is not of particular
importance. We have
\be
\mbox{Im} \; V(q,\omega)  
=g \; \om \; \frac{ D q^2 +r}{\om^2 + K_q^2} \;\; , \;\;
\ee
with $K_q=(D q^2+r)  \left(\xi^2 (q-q_0)^2 + a\right) $. We consider the limit 
$\ep \rightarrow 0$. The function $A(x)=\coth(x)+\tanh(-x)$ satisfies
$A(x\rightarrow 0)=1/x+O(x)$ and $A(|x|\gg 1) \rightarrow 0$. Hence
we evaluate the integral over $\omega$ as
\be
 \int_{-2T}^{2T} d\om \left( \frac{2T}{\om} \right) R(q,\omega) 
\frac{\om}{\om^2 + K_q^2} \simeq \frac{8 T^2}{K_q^2} R(q)
\;\; ,\;\;
\ee
where $R(q)=R(q,\omega \simeq T)$, and $2T < K_q$ is implied.

First we consider the case $q_0=0$. We see that the {\em dominant contribution} 
to the remaining integral $L$ over $q$ comes from {\em finite} 
$q>q_1= \sqrt{a} / \xi$ 
\bea
L=\int_{q_1}^{q_{max}} dq \; q^2 \; \frac{(D q^2+r) \; R(q)}{q \; K_q^2} 
\simeq  \;\; \;\; \\   \frac{R_0}{2B}  
\left \{ \frac{1}{a+\xi^2 q^2} 
+\frac{D}{2B} \ln\left( Dq^2+r \right)
-\frac{D}{2B} \ln\left( a+ \xi^2 q^2 \right)
\right \}_{q_1}^{q_{max}}  \;\; ,  \nonumber
\eea
with $q_{max}=1/2 \tau_o v_F$, $R_0=R(\bar q)$ and $B=aD-r\xi^2$. 

We consider in detail {\em  two different limiting cases}, 
namely $r=0$, $D> 0$ and $r > 0$, $D=0$.

For the case $r=0$, $D> 0$ we have
\be
L=\frac{R_0}{D} 
\left \{ \frac{1}{ a^2}  \ln \left( \frac{q_{max}}{q_1} \right)
+  \frac{1}{2 a } \left( \frac{1}{\xi^2 q_{max}^2 + a} 
- \frac{1}{\xi^2 q_{1}^2 + a } \right) \right \}  \; .
\ee
This yields $L \propto 1/a^2$. Then, for the case $r > 0$, $D=0$ we have
\be
L=-\frac{R_0}{2 \xi^2 r} 
\left \{ \frac{1}{\xi^2 q_{max}^2 + a} - \frac{1}{\xi^2 q_{1}^2 + a }  
\right \}  \;\; .\;\;
\ee
Here, $L \propto 1/(\xi^2 \; a)$. 

To estimate $R_0$ we consider the relation
$\arctan (x)-\arctan (y)=\arctan((x-y)/(1+x y))$, which yields
$R(q) \simeq \arctan \left( 2 k q/ m s_o \right)$.
The most relevant momenta are $k \simeq k_F$, thus yielding $k/m=v_F$.
Then we take $\bar q=q_{max}/2$ and we obtain
$R_0 \simeq \arctan(1) = \pi/4 $.

Next, we turn to the case of finite $q_0$ and, as above, 
we consider $q>q_1= \sqrt{a} / \xi$ and the limit $a \rightarrow 0$.

We have for all $r,D$
\begin{widetext}
\bea
L=\int_{q_1}^{q_{max}} dq \; \frac{q \; R(q)}{(Dq^2+r)(a+\xi^2 (q-q_0)^2)^2} 
\label{int3d} \\
\simeq \frac{ q_0 \;R_0 }{2 (Dq_0^2+r)\xi \; a}  \left \{ 
\frac{1}{\sqrt{a}} \arctan \left( \frac{\xi(q-q_0)}{\sqrt{a}}  \right)
+ \frac{\xi (q-q_0)}{a+\xi^2 (q-q_0)^2} 
+ O\left( \sqrt{a}  \right) \right \}_{q_1}^{q_{max}}  \;\; .\;\;
\nonumber
\eea
\end{widetext}

We now turn to 2-D. Again, 
$\theta$ is the angle between the vectors ${\bf k}$ and ${\bf q}$.
From the factor Im $G(k-q,\ep-\om)$, we have the angular integral
$I(q,\om)= \int_0^{2\pi} \frac{d \theta \; s_o}{s_o^2+(c+b \cos \theta)^2}$   
with $s_o,c,b$ as above.
For small $\omega$ and $k \simeq k_F$, $\ep \rightarrow 0$ we can make
the approximation 
$I(q,\om) \simeq \frac{2 \pi}{s_o} = I_0$.
As above, the integral over $\omega$ is
$\int_{-2T}^{2T} d\om \left( 2T/\om \right)  
(\om \; I_0)/(\om^2 + K_q^2) \simeq 8 T^2 \; I_0 /K_q^2 $.
Then, we have the remaining integral $L_2$ over $q$ in 2-D 
\be
L_2 = \int_{q_1}^{q_{max}} dq \; q \; \frac{I_0}{(Dq^2+r)(a+\xi^2 (q-q_0)^2)^2} 
\;\;.\;\;
\ee
We notice that the integrand is very similar to the 3-D case of
eq. (\ref{int3d}), and differing only in the factor $I_0$.
Hence, we obtain the same scaling of Im $\Sigma$ with $a$ as in 3-D.

In 3 dimensions the final result is
\be
\mbox{Im} \; \Sigma(k_F,\ep=0,T,a) = 
g /(2 \pi  v_F ) \; f_3(a,q_0) \; T^2 \;\;, 
\label{sig3d}
\ee
with $v_F$ the Fermi velocity. The function $f_3(a,q_0)$ is given in Table 1.
Therein $l_0=q_0/(Dq_0^2+r)$.

In 2 dimensions we have 
\be
\mbox{Im} \; \Sigma(k_F,\ep=0,T,a) =  (8 g \; \tau_o /\pi) 
\; f_2(a,q_0) \; T^2 \;\; . \;\;   \label{sig2d}
\ee
It turns out that $f_2(a,q_0) = f_3(a,q_0)$, 
so the prefactor dependence on $a$ is the {\em same} as in 3-D.
We note that the above dependence on $a$, $\xi$ and $T$ is valid 
for {\em all} $k$ of the order of $q_{max}$ or greater. For $k$ away from $k_F$
only the prefactors change.

A non-FL result Im $\Sigma \propto \sqrt{T}$ was obtained in \cite{maslov}
for {\em clean} metallic systems close to a FM QCP. We emphasize that herein 
we treat weakly disordered systems instead - c.f. the comment on 
$\tau_o^{-1}$ following eq. (\ref{gr0}) above.

\vspace{0.3cm} 
{\bf Calculation of the conductivity.}
We consider the {\em total} quasi-particle scattering rate
\be
\tau^{-1}(T,a) = \tau_{o,i}^{-1} + 2 \; \mbox{Im} \; \Sigma(k_F,\ep=0,T,a)= 2 S  
\;\;, \;\;   \label{sigm}
\ee
with $\tau_{o,i}^{-1}$ due to impurity scattering.

We calculate the total conductivity $\sigma$ from an infinite series 
of diagrams involving disorder and $V(q,\om)$. 
The $n$-th term of the conductivity series, shown in fig. 2, comprises 
$n$ impurity scattering
lines in parallel. We recall that two impurity lines
crossing each other introduce a small factor $1/\ep_F \tau_o \ll 1$ - c.f. 
refs. \cite{agd,lee}.
Hence we ommit any such diagram. In ref. \cite{gmr} we summed up
to infinite order another diagrammatic conductivity series, which includes
disorder and interactions (and yields experimentally observed positive
giant magnetoresistance).
 The Green's function here is taken as 
\be
G^{R,A}_*(k,\ep)=\frac{1}{\ep-\ep_k + \ep_F \pm i S} \;\; , \;\;  \label{grs}
\ee
i.e. it includes the self-energy of eq. (\ref{sigm})
due to the fluctuation potential $V(q,\om)$.
Due to momentum conservation, the momenta in the upper and lower lines of 
the $m$-th pair of $G^{R}_* G^{A}_*$ are the {\em same}.
The {\em major contribution} to these diagrams comes from assuming for 
the various vectors $k_1  = k_2 = k_3 = ... = k_n = k_{n+1} $.
Also $\ep=0$.

\begin{figure}[tb]
  \includegraphics[height=5.5truecm,width=6truecm]{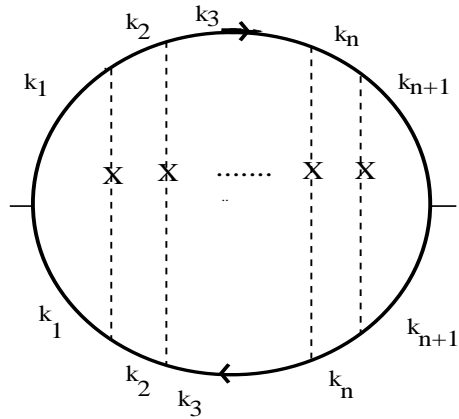}
\caption{Feynman diagram for the $n$-th term of the conductivity series. 
The dashed lines with crosses stand for impurity scattering. There are 
$n$ such impurity lines inside the conductivity bubble.
The Green's function here is given by eq. (\ref{grs}). }

\label{fig.2}
\end{figure}

Ommitting vertex corrections, the $n$-th term of this series is given by
($\sigma = \sum_{n=0}^{\infty} \sigma_n $)
\bea
\sigma_n = u_o^n \; \frac{2 e^2}{m^2} \; \int d{\bf k} \; k^2 \;
(G^R_*(k,0) G^A_*(k,0))^{n+1} \\
= u_o^n \; \frac{4 N_F e^2}{m} \;
\int_{-\ep_F}^{\infty} dx \frac{ \ep_F+x}{(x^2+S^2)^{n+1}}  \;\; , \;\;
\nonumber
\;\;
\eea
where $x=\ep_k - \ep_F$,
$N_F$ is the density of states at the Fermi level, $e$ is the charge of 
the electron, $m$ is the mass of the electron,
$u_o = n_{imp} V_{imp}^2$, with $n_{imp}$
the concentration of impurities, $V_{imp}$ the typical value of the impurity
scattering potential. Summing up this series,
and considering the energy scale $E_0 > \ep_F$, the result is
\bea
\sigma = \frac{ 4 e^2 N_F  }{m} 
\left\{ \frac{\ep_F}{\sqrt{S^2 - u_o}} \left[ \frac{\pi}{2}
+\arctan\left({\frac{\ep_F}{\sqrt{S^2 - u_o}}} \right) \right]  \right.
\nonumber  \\   \left.
+ \frac{1}{2} \ln\left(\frac{E_0^2 + S^2}{\ep_F^2 + S^2- u_o} \right) \right\} 
\;\; . \;\;    \label{totco} 
\eea
For $E_0 \sim O(\ep_F)$ and $\ep_F>S$ we may approximate 
$\sigma = 4 \pi \; e^2 N_F \ep_F / (m \; \sqrt{S^2 - u_o})$.

As the resistivity $\rho=1/\sigma$ comes out proportional to the scattering 
rate,
these results compare very favorably with data on CeCoIn$_5$ \cite{pag,bia},
Sr$_3$Ru$_2$O$_7$ \cite{gri} and YbRh$_2$Si$_2$ \cite{geg1,geg},
where $\rho$ displays typical FL $T^2$ dependence 
\be
\rho(T,a)=\rho_0 + A(a) \; T^2  \;\; , \;\;
\ee
with $\rho_0$=const., and
$A(a)$ diverging around the QCP, i.e. around $a=0$, as in our result.
With all the results of Table 1 in mind, we may concretely assume 
that the Gaussian 3-D case $q_0=0, r=0, D > 0$ applies
(with the alternative being the 3-D case $q_0=0, r>0, D = 0$, and $\xi$ 
{\em not} scaling with $a$, as mentioned below).  
Assuming the scaling $A(a) \propto a^{-2}$, as in the case
$q_0=0$ (FM) above, the diverging $A(a \rightarrow 0)$
in CeCoIn$_5$ yields $2s=1.29 \pm 0.1$ for the resistivity 
and $2s=1.34 \pm 0.1$ for the thermal resistivity \cite{bia,pag}.
Hence, for CeCoIn$_5$, the Wiedemann-Franz law, i.e. a constant ratio of 
the thermal conductivity $\kappa$ over the electrical conductivity $\sigma$,
times $T$, in the low $T$ limit, i.e. $\kappa/\sigma T $=const.,
is obeyed \cite{pag}. These experimental results are easily understood
in the frame of our calculation: conduction electrons carry both charge
and heat, while interacting via $V(q,\om)$ (only {\em small} energy
transfer is involved with this $V(q,\om)$).

The same scaling $A(a) \propto a^{-2}$ was also found to fit data in 
La$_{2-x}$Ce$_x$CuO$_4$ \cite{but} with $2s=0.38$, in overdoped
Tl$_2$Ba$_2$CuO$_{6+x}$  \cite{shib} with $2s=0.62$,
in CeAuSb$_2$ \cite{bali} with $2s=1.0$, and in YbAlB$_4$ \cite{naka}
with $2s=0.50$. 

In principle, {\em it should be possible} to probe the scattering rate 
through angle-resolved photoemission (ARPES) experiments (which require,
of course, that the materials in question be adequately cleavable).

\vspace{0.3cm} 
{\bf Calculation of the specific heat.}
We calculate the specific heat 
$C(T)= - T \partial^2 F(T)/\partial T^2 = \gamma \; T$, through 
the free energy $F(T)$ - c.f. fig 1(b) for the relevant Feynman diagram. 
To lowest order in $V(q,\om)$ we have
\bea
F(T) = T^2 \sum_{k,q,\ep_l,\om_m} V(q,i \om_m) \;  
G(k+q,i \ep_l+i \om_m) \; G(k,i \ep_l)  \nonumber  \\
= T \sum_{q,\om_m} V(q,i \om_m) \; \Pi(q,i \om_m) \;\; . \;\;
\eea
Here the Matsubara energies are $\ep_l=(2l+1)\pi T$ and $\om_m=2 m \pi T $,
$D_o = v_F^2  \tau_o / d$ ($d$=2,3 according to dimensionality) and \cite{lee}
$\Pi(q,i \om_m) = T \sum_{k,\ep_l} G(k+q,i \ep_l+i \om_m) \; G(k,i \ep_l)  
= N_F \left ( 1-\tau_o(D_o q^2 + |\om_m|) \right)$.

The integrand $ P(q,\om) =V(q,\om) \; \Pi(q,\om) $ has a branch cut for 
Im $\om=0$, which comes from Im $V(q,\om)$ 
(in which $-i \om \rightarrow +|\om_m|$). 
We ommit the $q,\om$ dependence of $\Pi(q,\om)$.
Then, via Cauchy's residue theorem, we obtain
$W = T \sum_{\om_m} P(q,i \om_m) = 
- \int_{-\infty}^{\infty} \frac{d\om}{\pi} \; n_B(\om) \; \mbox{Im} P(q,\om)$.
The Bose distribution function $n_B(\om) = 1/(e^{(\om/T)}-1)$ is commonly
approximated \cite{gkdef}
as $ n_B(\om) = (T/\om) \; \theta(T-|\om|) $. We first carry out
the $\om$ integration and then the integration over the momentum $q$ in
2 and 3 D.

Setting $Y_q=\xi^2 (q-q_0)^2 + a$ we obtain 
\bea
W \simeq -\int_{-T}^{T}  d\om \left(\frac{T}{\om}\right) \frac{\om}{(D q^2+r)} 
\; \frac{1}{Y_q^2 + \om^2/(Dq^2+r)^2}    \nonumber  \\
= -\frac{2 T}{Y_q}  \arctan \left( \frac{T}
{(D q^2+r) \; Y_q} \right) \;\; .\;\;
\eea

We consider the {\em low} $T$ regime 
$T < (D q^2+r) \; Y_q$  
and, as a result, $F(T)$ satisfies 
\be
F(T)=-\Gamma(T) \; T^2   \;\; . \;\; 
\ee

First, we consider the case $q_0=0$.
In 3-D we have, with $c_3=\frac{g \; N_F}{2 \pi^3} $
and $q_a=\min \left\{ 
\sqrt { T/(a D) }, \left( T/(D \xi^2) \right)^{1/4} \right\}$, 
the remaining integral over $q$

\begin{widetext}
\begin{equation}
\label{s.long}
\Gamma(T) = c_3 \; \int_{q_a}^{q_{max}}  \frac{dq \; q^2}{ (Dq^2+r) Y_q^2} 
= \frac{c_3}{2 B^2}
\left\{ \frac{q B}{a+\xi^2 q^2} 
-2\sqrt{r D} \arctan \left( \frac{q\sqrt{D}}{\sqrt{r}} \right)
+\frac{a D + r \xi^2}{\xi \sqrt{a}} 
\arctan \left(\frac{\xi q}{\sqrt{a}} \right) \right\}_{q_a}^{q_{max}}
\;\; .\;\;
\end{equation}
\end{widetext}

To proceed, we take the limit $T \rightarrow 0$ {\em first} and then the limit 
$a \rightarrow 0$, in order the extract the coefficient $\ga$.

Likewise, in 2-D, with $c_2= g \; N_F/ \pi^2$ we have for $q_0=0$
\bea
\Gamma(T) = c_2 \; \int_{q_a}^{q_{max}}  \frac{dq \; q}{ (Dq^2+r) Y_q^2} \\
= \frac{c_2}{2 B}
\left\{ \frac{1}{a+\xi^2 q^2} 
+\frac{1}{B} 
\ln \left(\frac{Dq^2+r}{a+\xi^2 q^2} \right) \right\}_{q_a}^{q_{max}}
\;\; .\;\; \nonumber
\eea

Next we turn to the case of {\em finite} $q_0$. We set $X_0=D q_0^2+r$.
Further, the minimum $q=q_{min}$ should now satisfy
$q_{min}=\sqrt{T/D} \; /(\xi \; \max\{q_0,q_{max}\})$.
In 3-D
\begin{widetext}
\begin{equation}
\Gamma(T) = c_3 \; \int_{q_{min}}^{q_{max}}  \frac{dq \; q^2}{ (Dq^2+r) Y_q^2} 
= \frac{c_3 \; q_0^2}{2 \xi \; X_0 \; a }  \left\{
\frac{1}{\sqrt{a}} \arctan\left(\frac{\xi(q-q_0)}{\sqrt{a}} \right)
+ \frac{\xi (q-q_0)}{a+\xi^2 (q-q_0)^2} 
+ O\left( \sqrt{a} \right) \right\}_{q_{min}}^{q_{max}}   \;\; .\;\; \label{gam3}
\end{equation}
\end{widetext}

And similarly for 2-D.

The coefficient $\ga$ in 3-D and 2-D is shown in Tables 1 and 2 
respectively. Due to phase space
considerations, for $q_0=0$ we have $\gamma_{2D} \propto \gamma_{3D}/\sqrt{a}$.

These results are consistent with CeCoIn$_5$ data \cite{bia} as a function 
of $a(H)$. They are also consistent with YbRh$_2$Si$_2$ \cite{geg1}
and Ge-doped YbRh$_2$Si$_2$ data \cite{geg}
(c.f. fig. 2 therein), with Sr$_3$Ru$_2$O$_7$ data \cite{rost},
and with CeRuSi$_2$ data \cite{flou}.

\vspace{0.3cm}
{\bf Kadowaki-Woods ratio.}
The scaling of the Kadowaki-Woods (KW) ratio $A/\gamma^2$ in 3-D and 2-D 
is shown in Tables 1 and 2 respectively.
Upon assuming the Gaussian regime,
c.f. eq. (\ref{gaus}), the KW ratio is constant {\em only} 
for the 3-D case $q_0=0, r=0, D>0$, modulo the logarithmic in $a$ divergence.
However, if, alternatively, $\xi$ and $a$ were independent parameters, and 
$\xi$ would {\em not} scale with $a$,
the KW ratio would be constant {\em only} for the 3-D case $q_0=0, r>0, D = 0$.
In all other cases either the ratio goes to {\em zero} for small $a$, or there
are {\em no} diverging prefactors $A,\ga$, contrary to the experiments.

Also we mention that substituting the factor $(D q^2 +r)$ in eq. (2) by
$(v_o \; q)$, i.e. the usual ferromagnetic damping with $v_o$ a constant, 
does {\em not} yield a (quasi-)constant KW ratio both in 3-D and 2-D.

An $a$-independent KW ratio
was observed in CeCoIn$_5$ \cite{bia}, YbRh$_2$Si$_2$ \cite{geg1},
Ge-doped YbRh$_2$Si$_2$ \cite{geg}, and YbAlB$_4$ \cite{naka}
though (in most cases) in a more restricted range
of $H$ than the scaling of the coefficient $A(a)$. 
E.g. in \cite{geg} the KW ratio increases as the control parameter
$a \rightarrow 0$, which is consistent with the logarithmic dependence
on $a$.

For the other 
materials mentioned above, the experimental data are incompletely known
with respect to the KW ratio.
Therefore, the possibility exists that they fall in some other 
case, among the ones mentioned in Tables 1 and 2, such that the
KW ratio is not constant.

\begin{table*}[ht]
\centering

\caption{ Table 1. Coefficients $f_3(a,q_0)$, $\ga$ and scaling of 
the Kadowaki-Woods ratio $A/\gamma^2$ ($A \propto f_3$) in 3-D 
as a function of the criticality parameter $a \ll 1$
and of the correlation length $\xi$. Here $c_3=g \; N_F/(2 \pi^3)$,
$B=a D - r \xi^2$, $q_{max}=1/2 \tau_o v_F$, 
$q_{min}=\sqrt{T/D} \; /(\xi \; \max\{q_0,q_{max}\})$, $q_1=\sqrt{a}/\xi$,
$l_0=q_0/(Dq_0^2+r)$, and $X_0=D q_0^2+r$.
Upon assuming the Gaussian regime,
c.f. eq. (\ref{gaus}), the Kadowaki-Woods ratio is constant for the case 
$q_0=0, r=0, D>0$, modulo the logarithmic in $a$ divergence.
In this case, the argument of the logarithm is $(q_{max} / a \sqrt{d_o} ) $.
C.f. text. } 

\begin{tabular}{|c|c|c|c|} \hline
  
 3-D   & $f_3(a,q_0)$ & $\ga$ & scaling of $A/\gamma^2 $  \\

\hline

$ q_0=0, r=0, D>0$ & $  \frac{1}{ a^2 \; D}\; \left [
\ln \left(\frac{ \xi q_{max}}{ \sqrt{a}} \right) -\frac{1}{4} \right]
$ & $ \frac{c_3 \; \pi}{2 \xi D \; a^{3/2} }$  & $ a \; \xi^2 \;
\ln \left( \xi \; q_{max} / \sqrt{a}  \right) $  \\

\hline

$ q_0=0, r>0, D=0$ & $ \frac{1}{ 4 \xi^2 r \; a} $ & 
$ \frac{c_3 \; \pi  r \xi }{2 B^2 \sqrt{a}}$ 
& constant  $ \xi^4$ \\

\hline

$ q_0>q_{max}>q_1>0 $ & $ \frac{l_0}{ a \; \xi^2}\left\{ \frac{1}{q_{max}-q_0}
-\frac{1}{q_1-q_0} \right\}
 $& 
$\frac{c_3 \; q_0}{ X_0 \; \xi^2 \; a} \;  \left( \frac{1}{q_{max}-q_0}  
-  \frac{1}{q_{min}-q_0}  \right) $ & $ a \;\xi^2 $ \\

\hline

$ q_{max}>q_0>q_1>0  $ & $ \frac{l_0 \; \pi}{ 2 \xi \; a^{3/2}}
 $ & 
$ \frac{c_3 \; q_0^2  \; \pi }{ X_0 \; \xi \; a^{3/2} } $ & 
$ a^{3/2} \; \xi $  \\

\hline

\end{tabular}



\end{table*}

\begin{table*}[ht]
\centering

\caption{ Table 2. Coefficients $f_2(a,q_0)$, $\ga$ and scaling of the 
Kadowaki-Woods ratio $A/\gamma^2$ ($A \propto f_2$) 
in 2-D as a function of the criticality parameter $a \ll 1$
and of the correlation length $\xi$. Here $c_2= g \; N_F/ \pi^2 $.
C.f. caption of Table 1 and text. } 

\begin{tabular}{|c|c|c|c|} \hline
  
  2-D   & $f_2(a,q_0)$ & $\ga$ &  scaling of $ A/\gamma^2 $    \\

\hline

$ q_0=0, r=0, D>0$ & $ \frac{1}{ a^2 \; D}\; \left [
\ln \left(\frac{ \xi q_{max}}{ \sqrt{a}} \right) -\frac{1}{4} \right]
$ & $ \frac{c_2 }{ D \; a } 
\left\{ \frac{1}{\xi^2  q_{max}^2}-\frac{1}{a} \right\} $ 
& $ a^2 \; \ln \left( \xi \; q_{max} / \sqrt{a} \right)$     \\

\hline

$ q_0=0, r>0, D=0$ & $ \frac{1}{ 4 \xi^2 r \; a} $ 
& $  \frac{c_2 }{ 2 r \xi^2 } \left\{ \frac{1}{a} - \frac{1}{r \xi} 
\ln\left( \frac{a}{\xi^2  q_{max}^2} \right) \right\}$ & $ a \;\xi^2 $ \\

\hline

$ q_0>q_{max}>q_1>0 $ & $ \frac{l_0}{ a \; \xi^2}\left\{ \frac{1}{q_{max}-q_0}
-\frac{1}{q_1-q_0} \right\} $ & 
$ \frac{c_2 }{\xi^2 \; a} \;  \left( \frac{1}{q_{max}-q_0}
- \frac{1}{q_{min}-q_0}  \right) $ 
& $ a \;\xi^2$ \\

\hline

$ q_{max}>q_0>q_1>0  $ & $ \frac{l_0 \; \pi}{ 2 \xi \; a^{3/2}} $ & 
$ \frac{c_2 \; \pi }{  \xi \; a^{3/2} } $ & 
$ a^{3/2} \; \xi$  \\

\hline

\end{tabular}

\end{table*}

In ref. \cite{millis}, using a different approach, diverging FL prefactors 
were obtained both for the resistivity and the specific heat. However,
the results differ from ours. Therein, the KW ratio is constant only for 2-D
FM fluctuations.

A number of experiments, probing quantities other than the above 
mentioned, suggest AFM behavior in CeCoIn$_5$ \cite{pag,kout}. 
A possible explanation is that both FM and
AFM fluctuations {\em coexist} in this material, with $A(a)$ and $\gamma(a)$
being determined {\em dominantly} by FM fluctuations.

We also consider an interaction with peaks 
at specific wavevectors $\vec{q}_{0i}$ (AFM case)
$V_*(q,\om)=\sum_{i=1}^{n_d} 
g/\{-i \om /(D q^2+r) +  \xi^2 (\vec{q}-\vec{q}_{0i})^2 + a\}$.
In 2-D for tetragonal symmetry $n_d=4$ and in 3-D for cubic symmetry $n_d=6$.
For small $a$, in 2-D and 3-D the potential $V_*(q,\om)$ 
gives the same scaling of the prefactors as for the finite $q_0$ cases
above.

\vspace{0.3cm}
{\bf Overview.}
In all, a consistent Fermi liquid description emerges from these
calculations. The renormalization of the fermions due to $V(q,\om)$ leads
back to the FL fixed point in a low-$T$ part of the phase diagram.
This is consistent with experiments
- e.g. c.f. fig. 3 of ref. \cite{pag} for the case $T^2$.
The prefactors for the scattering rate, the resistivity and the specific
heat diverge as power laws of the criticality parameter $a$.
According to our calculations, the Kadowaki-Woods ratio is constant only
for 3-D FM ($q_0=0$) fluctuations (possibly times the logarithmic 
in $a$ divergence).

We did not calculate explicitly the effective mass $m_*$ of the electrons. 
Experiments in Sr$_3$Ru$_2$O$_7$ \cite{merc} have indicated the absence 
of a magnetic field $H$ dependent renormalization of $m_*$ (definitely so 
for five out of the six bands). 
This result is not inconsistent with our calculations, where the diverging
overall prefactors, as a function of $a$, come from the small $q$ 
dependence of the potential $V(q,\om)$, and yield a uniform $a$ dependence
within the Fermi surface (c.f. the comment after eq. (\ref{sig2d})).

Finally, we comment on the linear in $T$ resistivity $\rho$ displayed by 
Sr$_3$Ru$_2$O$_7$ \cite{gri} and CeCoIn$_5$ \cite{pag} in the vicinity of 
the QCP and for not very low $T$. In ref. \cite{gkc} we developed a
fully microscopic FL model with a strong van Hove singularity (or, strong
peak in the density of states), located
at a characteristic energy $\ep_{vH}$ close to the Fermi level $\mu$.
This yields a quasi-particle scattering rate which is linear in $T$ 
for $T>(\mu-\ep_{vH})/4$. The model works very well for the cuprates, and 
this is how the linear in $T$ resistivity of La$_{2-x}$Ce$_x$CuO$_4$ \cite{but}
can be understood, given the small difference $\mu-\ep_{vH}$ for many
cuprates \cite{lu}, as shown by ARPES expts. Such expts. \cite{tama}   
on Sr$_3$Ru$_2$O$_7$ 
indeed yielded $\mu-\ep_{vH}=4$ meV, in agreement with our 
model \cite{gkc}. It is fair to attribute the linear in $T$ resistivity
of CeCoIn$_5$ to the same mechanism, i.e. originating from a van Hove 
singularity, which resides close to the Fermi surface in the vicinity 
of the QCP.
Moreover, it is reasonable to view the regime $\rho \propto T^b$, $1<b<2$,
displayed by Sr$_3$Ru$_2$O$_7$ \cite{gri} and CeCoIn$_5$ \cite{pag},
as a smooth {\em transient} between $b=1$ and $b=2$.

\vspace{.3cm}
$^*$ e-mail : kast@iesl.forth.gr ; giwkast@gmail.com

\newpage


\vspace{0.7cm}

\centerline{\bf Supplementary Information}

\vspace{0.6cm}
{\bf Appendix A : On the effective interaction $V(q,\om)$}
\vspace{0.3cm}

$V(q,\om)$ of eq. (2) can only be derived in the context of
an RPA-type approach, in the spirit of references [15-20] cited
in the article. This fact is also emphasized in the recent article
by Y. Wang and A.V. Chubukov in Phys. Rev. B {\bf 92}, 125108 (2015).

\vspace{0.6cm}
{\bf Appendix B : On the calculation of the scattering rate}
\vspace{0.3cm}

In the limit $T \rightarrow 0$ the thermal function 
$X= \coth(\om/2T) \; + \; \tanh((\ep-\om)/2T)$ in eq. (4) becomes
$X=2$ for $2T < \om <\ep$, and $X=0$ for $\om<-2T$ and $\om>\ep$.
Then the integration over $\om$ - compare with eq. (7) - amounts to
\bea
2 \int_{2T}^{\ep} d\om \; \text{Im} \; V(q,\om) \; R(q,\om)  \nonumber \\
\simeq g \ R_0 \; \ln\left( \frac{K_q^2+\ep^2}{K_q^2+ 4T^2} \right)
\simeq g \; R_0 \; \frac{\ep^2}{K_q^2} \;\; , \;\;
\eea
for $K_q > \ep$. The rest of the algebra proceeds as in eq. (8) and
onwards. Thus the scattering rate scales like $\ep^2$ as well, 
as expected for the FL regime.

\vspace{0.3cm}
We also give some details of the 2-D calculation in the main text.
The angular integral $I(q,\om)$ is given by
\bea
I(q,\om)= \int_0^{2\pi} \frac{d \theta \; s_o}{s_o^2+(c+b \cos \theta)^2} 
\nonumber \\
=\pi \; \left( \frac{1}{\sqrt{ b^2-c^2-2i s_o c+s_o^2}} 
+ \text{c.c.}  \right)   \;\;,\;\;
\eea
with $c,b$ as above.
Then, taking
\be
\al = b^2-c^2+s_o^2 >0, \;\;,\;\; \beta = 2 s_o c\;\;,\;\;
\al \gg |\beta|  \;\;,\;\;
\ee
we can approximate
\be
I(q,\om) \simeq  2 \pi\; 
\left( \frac{\al}{\al^2+\beta^2} \right) ^{1/2} 
\simeq \frac{2 \pi}{s_o}   \;\;. \;\;
\ee

\vspace{0.6cm}
{\bf Appendix C : On the calculation of the infinite series for the 
conductivity}
\vspace{0.3cm}

Setting $Z_0=4 N_F e^2/m$, the n-th term of the conductivity series is given by
\bea    
\De_n =u_o^n \; Z_0 \int_{-\ep_F}^{\infty} dx \frac{ \ep_F+x}{(x^2+S^2)^{n+1}} 
\nonumber \\
= u_o^n \; Z_0 \; ( \ep_F \; \Ga_n - \ep_F \; A_n + B_n) \;\; , \;\;
\;\;
\eea
where $x=\ep_k - \ep_F$ and
\bea
\Ga_n = \int_{-\infty}^{\infty} \frac{dx }{(x^2+S^2)^{n+1}} = 
\frac{\pi}{2^n \; S^{2n+1}} \; \frac{(2n-1)!!}{n!}
\;\;, \\
A_n = \int_{-\infty}^{-\ep_F} \frac{dx}{(x^2+S^2)^{n+1}} \;\;, \\
B_n = \int_{-\ep_F}^{\infty} \frac{dx \; x}{(x^2+S^2)^{n+1}} \;\;.
\eea

We easily obtain
\be
\Ga = \sum_{n=0}^{\infty} u_o^n \; \Ga_n 
= \frac{ \pi}{\sqrt{S^2 - u_o}}
\;\; ,\;\;  
\ee
\be
A = \sum_{n=0}^{\infty} u_o^n \; A_n = 
\frac{1}{\sqrt{S^2 - u_o}} 
\left[ \frac{\pi}{2} - 
\arctan\left( \frac{\ep_F}{\sqrt{S^2 - u_o}}\right) \right]
\;\; .\;\; 
\ee
For $n=0$ {\em only}, we consider as the upper limit of integration the energy
scale $E_0>\ep_F$, instead of infinity (this integral is ultra-violet
divergent), and we obtain
\be
B_0 = \frac{1}{2} \int_{\ep_F^2}^{E_0^2} \frac{dy}{y+S^2} = 
\frac{1}{2} \ln\left(\frac{E_0^2+S^2}{\ep_F^2+S^2}\right) \;\;. \;\;
\ee
For $n \ge 1$ we take infinity as the upper limit of integration,
thus obtaining
\be
\sum_{n=1}^{\infty} u_o^n \; B_n = \frac{1}{2} 
\ln\left(\frac{\ep_F^2+S^2} {\ep_F^2+S^2- u_o}\right) \;\;. \;\;
\ee
Hence
\be
B=\sum_{n=0}^{\infty} u_o^n \; B_n = \frac{1}{2} 
\ln\left(\frac{E_0^2+S^2} {\ep_F^2+S^2- u_o}\right) \;\;. \;\;
\ee

Putting all these terms together yields the total conductivity given
by eq. (\ref{totco}) above.


\begin{thebibliography}{99}


\bibitem{revi}
v. L{\"o}hneysen H., Rosch A., Vojta M. and W{\"o}lfle P.,
Rev. Mod. Phys. {\bf 79}, (2007) 1015.

\bibitem{pag}
Paglione J. et al., Phys. Rev. Lett. {\bf 97}, (2006) 106606.

\bibitem{bia}
Bianchi A. et al., Phys. Rev. Lett. {\bf 91}, 257001 (2003).

\bibitem{gri}
Grigera S.A. et al., Science {\bf 294}, 329 (2001).

\bibitem{geg1}
Gegenwart P. et al., Phys. Rev. Lett. {\bf 89}, 056402 (2002).

\bibitem{geg}
Gegenwart P. et al., Phys. Rev. Lett. {\bf 94}, 076402 (2005).

\bibitem{but}
Butch N.P. et al., PNAS {\bf 109}, 8440 (2012).

\bibitem{shib}
Shibauchi T. et al., PNAS {\bf 105}, 7120 (2008). 

\bibitem{bali}
Balicas L. et al., Phys. Rev. B {\bf 72}, 064422 (2005).

\bibitem{naka}
Nakatsuji S. et al., Nature Phys. {\bf 4}, 603 (2008). 

\bibitem{rost}
Rost A.W. et. al., Science {\bf 325}, 160 (2009).

\bibitem{flou}
Flouquet J. et al., Physica B {\bf 319}, 251 (2002).

\bibitem{agd}
Abrikosov A. A., Gorkov L. P. and  Dzyaloshinski I. E.,
Methods of Quantum Field Theory in Statistical Physics,
Prentice-Hall (Cliffwoods, NY),
(1964).

\bibitem{lee}
Lee P.A. and Ramakrishnan T.V., Rev. Mod. Phys. {\bf 57}, (1985) 287.

\bibitem{her}
Hertz J., Phys. Rev. B {\bf 14}, 1165 (1976).

\bibitem{mil1}
Millis A.J.,  Phys. Rev. B {\bf 48}, 7183 (1993).

\bibitem{millis}
Millis A.J., Schofield A.J., Lonzarich G.G. and Grigera S.A.,
Phys. Rev. Lett. {\bf 88}, 217204 (2002).

\bibitem{kim}
Kim Y.B. \and Millis A.J., Phys. Rev. B {\bf 67}, 085102 (2003).

\bibitem{chu}
Chubukov A.V., Galitski V.M. \and Yakovenko V.M., 
Phys. Rev. Lett. {\bf 94}, 046404 (2005).

\bibitem{maslov}
Maslov D.L. \and Chubukov A.V., Phys. Rev. B {\bf 79}, 075112 (2009). 

\bibitem{gkdef}
Kastrinakis G., Phys. Rev. B. {\bf 72}, 075137 (2005); in this work 
it was 
shown that the $T \rightarrow 0$ saturation of the electron dephasing rate,
observed in numerous expts., can be attributed to the elastic scattering 
from magnetic impurities.

\bibitem{gmr}
Kastrinakis G., Europhys. Lett. {\bf 42}, 345 (1998).

\bibitem{kout}
Koutroulakis G. et al., Phys. Rev. Lett. {\bf 104}, 087001 (2010).

\bibitem{merc}
Mercure J.-F. et al., Phys. Rev. B {\bf 81}, 235103 (2010).

\bibitem{gkc}
Kastrinakis G., Physica C, {\bf 340}, 119 (2000); 
Kastrinakis G., Phys. Rev. B. {\bf 71}, 014520 (2005).

\bibitem{lu}
Lu D.H. et al., Phys. Rev. Lett. {\bf 76}, 4845 (1996).

\bibitem{tama}
Tamai A. et al., Phys. Rev. Lett. {\bf 101}, 026407 (2008).

\end{thebibliography}
\end{document}